\documentclass[twocolumn,preprintnumbers,amsmath,amssymb,pra]{revtex4}

\usepackage{graphicx,amssymb,color}
\usepackage{dcolumn}% Align table columns on decimal point
\usepackage{bm}% bold math

\newcommand{\bfr}{{\bf r}}

\begin{document}

\title{Thermal activation of vortex-antivortex pairs in quasi-2D Bose-Einstein condensates}
\author{T.~P. Simula}
%\email{tapio@physics.otago.ac.nz}
\author{P.~B. Blakie}
\affiliation{Department of Physics, University of Otago, PO Box 56, Dunedin, New Zealand}

%\keywords{Bose-Einstein condensation, superfluidity, Berezinskii-Kosterlitz-Thouless phase transition, vortex}
%Use showkeys class option if keyword display desired

\begin{abstract}
Here we show, by performing \emph{ab initio} classical field simulations that two distinct superfluid phases, separated by thermal vortex-antivortex pair creation, exist in experimentally producible quasi-2D Bose gas. These results resolve the debate on the nature of the low temperature phase(s) of a trapped interacting 2D Bose gas.
\end{abstract}

\maketitle

% Second intro
The phenomena of superconductivity and superfluidity are striking manifestations of the r\^ole played by quantum statistics at low temperatures. Altering the temperature or effective dimensionality may radically change the physical properties of quantum degenerate systems. A well known consequence of this is that, unlike in 3D, there is no Bose-Einstein condensation (BEC) for a homogeneous 2D ideal-gas in the thermodynamic limit at any finite temperature \cite{Mermin1966a,Hohenberg1967a}. Nevertheless, the Berezinskii-Kosterlitz-Thouless (BKT) vortex binding-unbinding phase transition allows the emergence of superfluidity in 2D systems \cite{Berezinskii1971a,Kosterlitz1973a,Minnhagen1987a}. This superfluid transition has been experimentally observed in liquid helium thin films \cite{Bishop1978a}, superconducting Josephson-junction arrays \cite{Resnick1981a}, and in spin-polarized atomic hydrogen \cite{Safonov1998a}. Although weak particle interactions alone are not sufficient to change the situation, an external confinement modifies the density of states in such manner that the critical point of BEC is elevated to finite temperatures \cite{Bagnato1991a}. Therefore it is not certain \emph{a priori} whether the transformation from normal to superfluid in such systems is BEC or BKT-type transition.

Significant advances in trapping ultra-cold atoms, and in particular the use of optical lattices, has made production of quasi-2D quantum gases experimentally feasible \cite{Gorlitz2001a,Schweikhard2004a,Rychtarik2004a,Smith2005a,Stock2005a}. Low temperature ordering in these systems leads to the competition between BEC and BKT-type transitions. Long wave-length phase fluctuations have been predicted to destroy the phase coherence of a pure condensate at low temperatures leaving the system in a state of so-called quasi-condensate, however, the structural details of such a state have remained under speculation \cite{Simula2005a,Popov1983a,Petrov2000a,Prokofiev2002a,Andersen2002a,Gies2004a,Holzmann2005a,Trombettoni2005a}. In this Letter we provide quantitative evidence for the existence of a vortex-free condensate phase and a BKT-type phase manifested by the presence of vortex-antivortex pairs, using a formalism that allows direct observation of the transition between these two distinct superfluid phases. Furthermore, we show that the phase defects observed in the experiment by Stock \emph{et al.} \cite{Stock2005a} are readily explained in terms of spontaneous activation of vortex-antivortex pairs in the temperature regime of the predicted BKT-type phase.

% Physics background
The statistical probability, $p$, for the excitation of a vortex-antivortex pair in a phase coherent Bose-Einstein condensate is proportional to the Boltzmann factor, $e^{-F/k_{\rm B}T}$, where $F$ is the free energy cost of such topological excitation, $k_{\rm B}$ is the Boltzmann constant and $T$ is the temperature. The condition $p=1$ may be used to estimate the critical temperature, $T_c$, separating the vortex-free and vortex-pair phases. Considering a single pair in the vicinity of the 2D harmonic trap centre, the hydrodynamic approximation yields \cite{Simula2005a}
\begin{equation}
\frac{T_c}{T_0}=\sqrt{\frac{\beta}{\beta+g\ln^2(8gmN_0/\hbar^2)}},
\label{Eq1}
\end{equation}
where, $\beta=2\pi^3\hbar^2(\ln 2+\epsilon)^2/3m$, $N_0$ is the number of particles of mass $m$ in the condensate phase, $g$ is the strength of the particle interactions, and $T_0$ is the critical temperature for an ideal-gas to undergo Bose-Einstein condensation. The constant $\epsilon$ is the core energy per particle of a vortex pair, expressed in the trap units. 

In order to numerically study the low temperature phase diagram of a trapped quasi-2D Bose gas and the validity of Eq.~(\ref{Eq1}), we divide the system into classical and so-called incoherent regions depending on the mode occupation of the single particle eigenstates. The classical region consists of the highly occupied modes that are described by the projected Gross-Pitaevskii equation  
\begin{equation}
i\hbar\partial_t\Psi= -\frac{\hbar^2}{2m}\nabla^2\Psi  +V_{\rm ext}\Psi+g\mathcal{P}\{|\Psi|^2\Psi\},
\label{Eq2}
\end{equation}
where $V_{\rm ext}$ denotes an external potential and the projector $\mathcal{P}$ serves to restrict evolution of the classical field, $\Psi(\bfr,t)$, to within its subspace \cite{Blakie2005a,Davis2005a}. The incoherent region consists of the modes of low occupation which are treated using the semi-classical Hartree-Fock approximation. The classical and incoherent regions are taken to be in equilibrium with each other. The former is most conveniently represented using the harmonic oscillator eigenstates with energies below certain cutoff energy, $E_{\rm{cl}}$. This energy cutoff needs to be chosen such that the mean occupation of all the modes in the classical region is $\gtrsim 1$. The normalization of $\Psi(\bfr,t)$ determines the number of particles $N_{\rm cl}>N_0$ in the classical region. To simulate the experiment \cite{Stock2005a}, we concentrate on a single quasi-2D site of their optical lattice and perform simulations employing a 3D harmonic trap with respective radial and axial frequencies $\{\omega_\perp,\omega_z\}=2\pi\times \{106,4000\}$ Hz. This choice corresponds to the shorter lattice period in the experiment. The time and spatial scales in our calculations are $t_0=2\pi/\omega_\perp$ and $a_0=\sqrt{\hbar/2m\omega_\perp}$, respectively. Depending on the total particle number, $N$, the energy cutoff is in the range, $E_{\rm cl} = (32-56)\;\hbar\omega_\perp$, resulting in the non-perturbative treatment of several hundred lowest modes of the many-body system. 

\begin{figure}[!th]
\center
\includegraphics[width=86mm,viewport=120 220 470 550,clip]{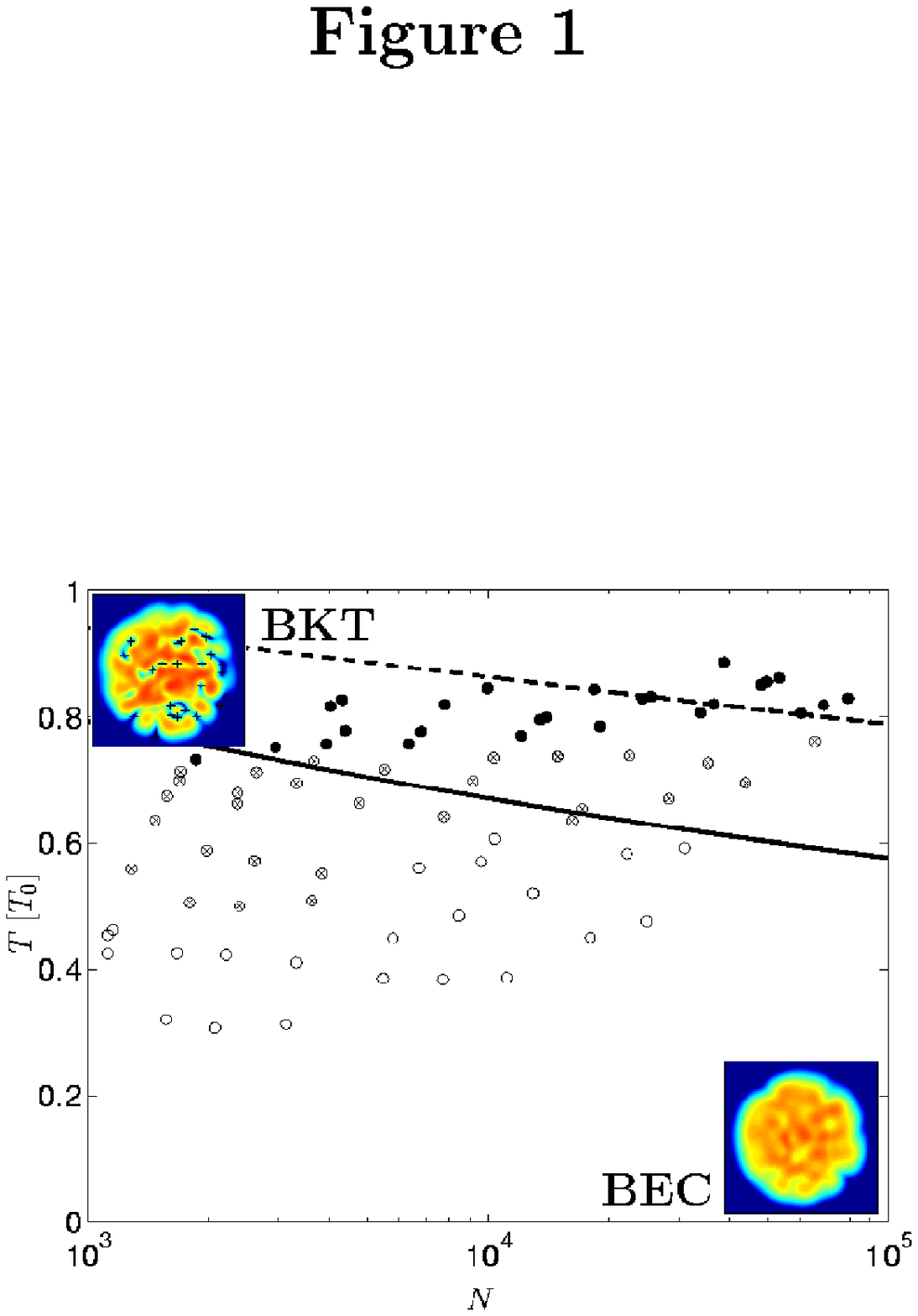}
\caption{Low temperature phase diagram for quasi-2D Bose gas. The markers denote different vortex-pair excitation probabilities $p=0$ ($\circ$), $0 < p < 0.25$ (\protect\raisebox{0.3ex}{\tiny{$\otimes$}}) and $p>0.25$ ($\bullet$) inside a circular region whose diameter equals the Thomas-Fermi radius. The lines are analytical estimates using Eq.~(\ref{Eq1}) with $\epsilon=0$ \mbox{(-\hspace{-.3mm}-\hspace{-.3mm}-)} and $\epsilon=1/2$ (-\hspace{.5mm}-).}
\label{Fig1}
\end{figure}

% Simulation method
Each simulation proceeds as follows. The initial mode distribution of the classical field, $\Psi(\bfr,t)$, is first allowed to equilibrate for 10 trap periods, \emph{i.e.} $10\;t_0$, after which $N_0$, $N$ and the fractional temperature, $T/T_0$, are extracted by averaging over 500 microstates, sampled uniformly over another $10\;t_0$. The reference ideal-gas condensation temperature, $T_0$, is obtained by exact summation of the partition function for the quasi-2D harmonic oscillator. Subsequently we numerically detect vortex pairs by locating the phase singularities in the classical field, $\Psi(\bfr,t)$. To measure the vortex-pair excitation probability, $p$, we only include vortices observed inside a circle whose diameter is equal to the Thomas-Fermi radius for a pure condensate of $N_0$ particles. Thus we evaluate, $p$, by counting the average number of vortex pairs in the 500 microstates sampled for each simulation.

\begin{figure}[!th]
\center
\includegraphics[width=86mm,viewport=120 220 470 550,clip]{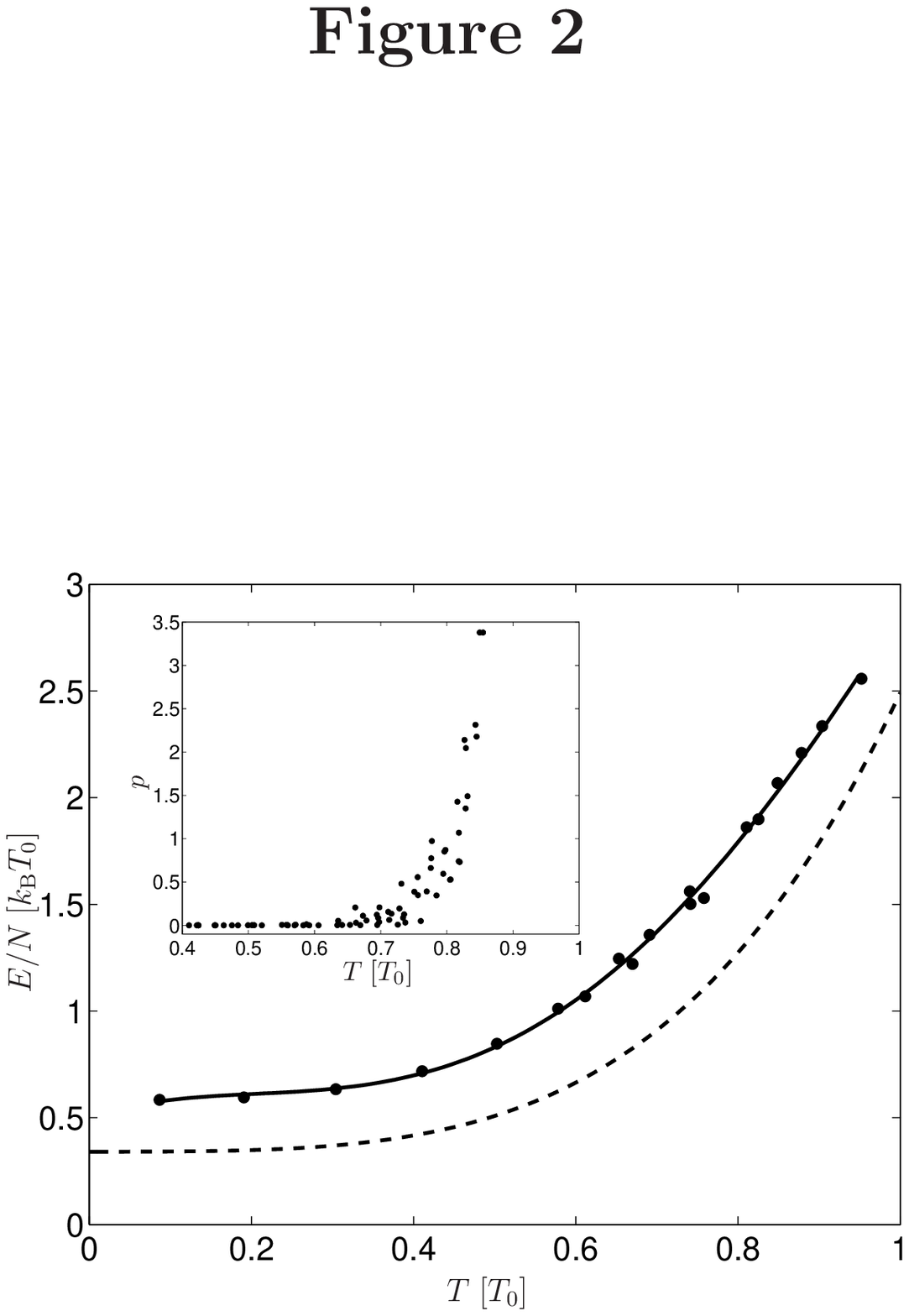}
\caption{Internal energy for $N=1\times 10^4$ particles (main frame) and the vortex excitation probability for a range of $N$ (inset) as functions of the fractional temperature. The solid line is a fourth order polynomial fit to the data. The dashed line is the ideal-gas result.}
\label{Fig2}
\end{figure}

The computed superfluid phase diagram for a trapped quasi-2D Bose gas is shown in Fig.~\ref{Fig1}. The insets show typical planar density distributions in the two superfluid phases. We have categorized the simulated systems according to the measured vortex-pair excitation probabilities $p=0$ ($\circ$), $0 < p < 0.25$ (\protect\raisebox{0.3ex}{\tiny{$\otimes$}}) and $p>0.25$ ($\bullet$). The lines are plotted using Eq.~(\ref{Eq1}) with (-\hspace{0.5mm}-) and without (-\hspace{-.3mm}-\hspace{-.3mm}-) the core contribution $\epsilon$. Two distinct phases---vortex-free and vortex-filled---are evident in the diagram. The analytic result provides a fair estimate of the critical temperature for the transition between the two superfluid phases. The temperature range in which phase defects were observed in the experiment is consistent with that of the BKT-type phase in Fig.~\ref{Fig1}. The slight upward tendency in the computed transition temperature with increasing $N$ is probably due to axial degrees of freedom becoming thermally excited, causing the system characteristics to approach three dimensionality.

% VORTEX EXCITATION PROBABILITY FIGSI
Figure \ref{Fig2} shows the internal energy per particle of the gas as a function of the fractional temperature for $N=10^4$ particles. The computed vortex-pair excitation probability, $p$, corresponding to the data in Fig.~\ref{Fig1} is displayed in the inset. The solid line in the main figure is a fourth order polynomial fit to the data. The interactions impose an overall increase in the internal energy in comparison to the ideal-gas result (-\hspace{0.5mm}-). This energy characteristic alone is too coarse to reveal latent heat at the vortex-pair creation transition in this inhomogeneous mesoscopic system. However, as is evident in the inset, the measure of vorticity provides a useful order parameter to characterize the transition.

At the lowest temperatures no vortices are present in the system. They first emerge in the low density regions of the cloud and are only able to nucleate closer to the trap centre as the temperature increases. We note that at intermediate temperatures the phase coherent central condensate and outer shell of phase fluctuating superfluid co-exist, while at even higher temperatures the whole cloud turns critical with vortex pairs able to form and annihilate everywhere. Ultimately, as temperature increases, superfluidity is lost when the system crosses the BKT-type transition to the normal state. Thus the external confining potential, which fundamentally allows BEC to exist in 2D, also leads to a spatial dependence of the critical temperature for the spontaneous vortex-pair creation. The lifetime of vortex-pairs is dependent on the local superfluid density, being shorter in the high density central regions.

%INTERFERENCE
In the experiment a set of quasi-2D systems were realized using an optical lattice, and were subsequently allowed to expand in free space and interfere with each other. The presence of phase defects were inferred from the distortion of the interference patterns. In order to fully explain those observations, we have simulated this interference procedure. We construct a 3D wavefunction corresponding to a lattice of quasi-2D systems,
where the individual components are taken as independent microstates from a single classical field simulation. The lattice parameters are chosen to match those used in the experiment. Subsequently, we propagate thus prepared wavefunction according to a trap-free time-dependent Gross-Pitaevskii equation. Upon release from the external potential, mean-field effects on the evolution rapidly decay and the momentum distribution freezes, allowing us to infer the asymptotic spatial interference patterns. In practise they are obtained by Fourier transforming the spatial wavefunctions after a $0.5\;t_0$ time-of-flight. 

\begin{figure}[!th]
\center
\includegraphics[width=86mm,viewport=120 300 470 620,clip]{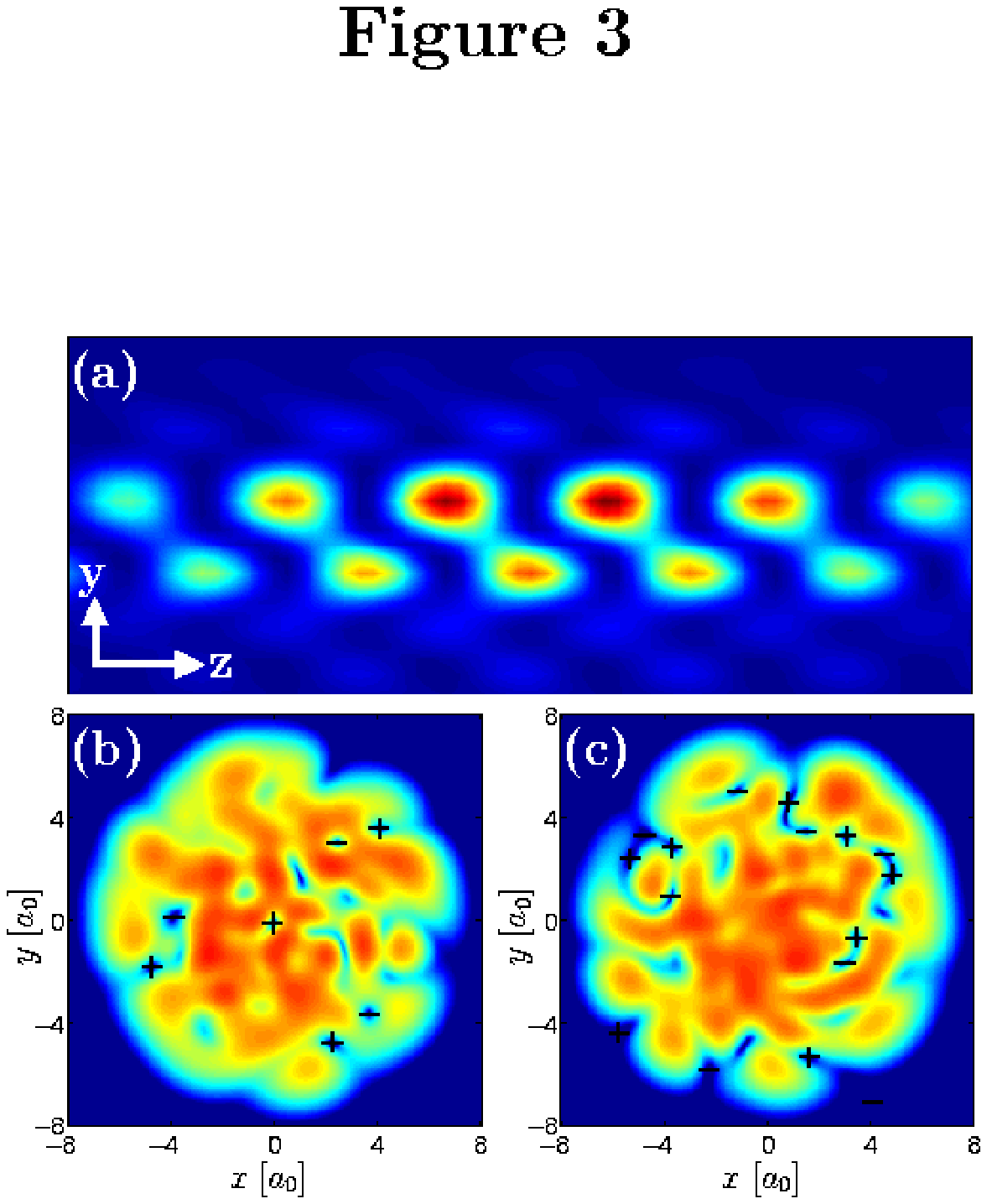}
\caption{Interference pattern (a) produced by two independent classical fields (b) and (c) at temperature $T=0.86\;T_0$. The relevant particle numbers are $N_{\rm cl}=3.0\times 10^3$ and $N=4.0\times 10^4$. The zipper structure in (a) is the telltale signature of the phase singularity associated with the central vortex in (b). The locations of vortices and antivortices are marked by $+$ and $-$ signs, respectively.}
\label{Fig3}
\end{figure}

Figure \ref{Fig3}(a) shows a result of such numerical interference experiment for the two classical fields whose planar density distributions are shown in Figs.~\ref{Fig3}(b) and (c). The characteristic zipper pattern in the interference is predominantly caused by the phase singularity associated with the isolated vortex in the centre of Fig.~\ref{Fig3}(b). Despite the transient imbalance in the number of oppositely charged vortices in Fig.~\ref{Fig3}(b), the value of angular momentum remains zero by virtue of the spatial vortex distribution. At low temperatures only straight fringes are produced in the interference patterns, manifesting the global phase coherence of the condensates. In the presence of vortex pairs, zippers and more subtle structures emerge revealing the underlying phase fluctuations.

We have verified that the zipper pattern observed in the experiments is indeed produced by an isolated vortex and that closely bound vortex-antivortex pairs do not significantly alter the interference pattern. However, such isolated vortices are remnants of vortex pairs which have dissociated, since conservation of angular momentum should prevent formation of unpaired vortices. In the vicinity of the vortex-pair creation transition only tightly bound vortex pairs exist but at higher temperatures, deeper in the BKT-type phase, seemingly independent vortices are abundant in the system since the energy cost of unbinding an existing pair is comparable to the cost of creating a new pair. We have verified such dynamic behaviour by visually observing the creation of vortex pairs and their subsequent dissociation and/or annihilation in simulations. The isolated vortices are most clearly identified by the interference pattern they produce, whereas the lack of resolution in this type of detection method obscures the observation of tightly bound vortex pairs. This, together with the relatively narrow temperature range of the BKT-type phase for these system parameters, could explain the rather low probability, $\approx 10\%$, for observing a clear signature of phase defects in the experiments. Therefore, further development of detection methods may be required to observe the onset of the vortex-pair creation (\emph{i.e.} BEC-BKT) transition. The probability for detecting phase defects could perhaps be enhanced by imparting angular momentum to the system to artificially separate the paired vortices.

Vortex-antivortex pairs are the only phase defects observed in the simulations. Other possibilities, such as dark solitons, can be excluded from the picture since their cost of energy, $F$, is much greater than that of a vortex(pair). And even if excited, oppositely travelling solitons would rapidly decay into vortices via the snake instability. However, at very high temperatures the vortices are occasionally seen to arrange themselves into a chain and percolate throughout the cloud in anticipation of the transition to a normal state. 

%DISCUSSION
To conclude, we have performed \emph{ab initio} classical field simulations of quasi-2D Bose-fields and have characterized the low temperature phases for such systems over a wide parameter range. We have provided strong evidence supporting the view that the BKT-type phase was observed in the recent experiment by Stock \emph{et al.} \cite{Stock2005a}. Moreover, we have demonstrated the limited sensitivity of the experimental detection method, based on pairwise interference of individual clouds, to tightly bound vortex-antivortex pairs. Our results further vindicate the applicability of classical field methods to the study of ultra-cold atomic gases. An additional benefit of our approach is that it allows for a direct study of the proliferation of vortices in the normal to superfluid (BKT-type) phase transition.

\begin{acknowledgments}
Financial support from the Academy of Finland and Otago University Research Grant are acknowledged.
\end{acknowledgments}

\end{document}